# Data for the green synthesis of exceptional braided, helical carbon nanotubes and nano spiral platelets made directly from $CO_2$

*Xinye Liu,[1] Gad Licht,[2] Stuart Licht[1,]**

[1]X. Liu, Prof. S. Licht*
Dept. of Chemistry, George Washington University, Washington DC 20052, USA.
E-mail: slicht@gwu.edu

[2]G. Licht, C2CNT Corp, 1035 26 St NE, Calgary, AB T2A 6K8, Canada.

**Abstract**

This article describes base-line data used in conjunction with the *Materials Today Chemistry* article "The green synthesis of exceptional braided, helical carbon nanotubes and nano spiral platelets made directly from $CO_2$." The data includes Raman, XRD, SEM, TEM, EDX and high contrast TEM of conventional carbon nanotubes (CNTs) synthesized from carbon dioxide electrolytic splitting in molten electrolyte for comparison to synthesized helical carbon nanotubes grown in the main article. This data article also presents SEM showing that uncontrolled electrolytic $CO_2$ splitting leads to bent and deformed CNTs, rather than straight or helical CNTs, and SEM that helical carbon nanotubes and helical carbon nanoplatelets can be produced with only partial control of the electrolysis current and with no defect inducing agents. A novel, cheap synthesis of high purity helical nanocarbon is discovered. A baseline and comparative experimental data for non-helical analogues is presented. Both the non-helical and helical nanocarbon are synthesized from the greenhouse gas $CO_2$ by molten carbonate electrolysis.

An occasionally reported unusual, coiled allotrope of carbon nanomaterial morphology is termed coiled or Helical Carbon NanoTubes (HCNTs) [1-20], which had been synthesized by CVD. In the main article of this series, HCNTs are instead synthesized from $CO_2$ by electrolysis in molten carbonates. The Licht C2CNT group has introduced a novel chemistry, which turns $CO_2$ into Carbon NanoTubes (CNTs), and other carbon nanomaterials and $O_2$ [20-35]. The Carbon dioxide to CNTs (C2CNT) process offers a high-efficiency method for production to directly remove $CO_2$ from the atmosphere or chemical/energy plants and turn it into useful products.

Dissolution:  $CO_2(gas) + Li_2O(soluble) \rightarrow Li_2CO_3(molten)$  (1)

Electrolysis:  $Li_2CO_3(molten) \rightarrow C(CNT) + Li_2O(soluble) + O_2(gas)$  (2)

Net:  $CO_2(gas) \rightarrow C(CNT) + O_2(gas)$  (3)

An important component of the C2CNT growth process is transition metal nucleated growth, such as the addition of Ni powder which leads to clearly observable CNT walls as shown in Figure S1. In the figure, TEM shows the CNT walls, and that the graphene spacing between the CNT walls is the expected of 0.34nm for graphene layers.

Experimental design, materials and methods: Lithium carbonate ($Li_2CO_3$, 99.5%), lithium oxide ($Li_2O$, 99.5%), and ferric oxide ($Fe_2O_3$, 99.9% metals basis) are combined to form various molten electrolytes.

Electrolyses are driven at a constant current density as described. Smaller area electrolyses ($\leq$ 5cm$^2$) are conducted in an alumina crucible and larger in a stainless steel 304 case. Nickel, Inconel 718, or Nichrome, was the (oxygen generating) anode as indicated in the text, and Monel or Muntz brass as the cathode also as indicated in the text. During electrolysis, the carbon product accumulates at the cathode, which is subsequently removed and cooled. After electrolysis the product remains on the cathode but falls off or peels off when the cathode is extracted, cooled, and tapped. The product is washed with either DI water or up to 6m HCl (both yield similar product, but the latter solution accelerates washing), and separated from the washing solution by either paper filtration or centrifugation (both yield similar product, but the latter accelerates the separation).

*Characterization:* The carbon product was washed, and analyzed by PHENOM Pro Pro-X SEM (with EDX), FEI Teneo LV SEM, and by FEI Teneo Talos F200X TEM (with EDX). XRD powder diffraction analyses were conducted with a Rigaku D=Max 2200 XRD diffractometer and analyzed with the Jade software package. Raman spectra were collected with a LabRAM HR800 Raman microscope (HORIBA). This Raman spectrometer/microscope uses an incident laser light with a high resolution of $0.6 cm^{-1}$ at a 532.14nm wavelength. XRD was measured with a Rigaku D=Max 2200 XRD.

Our previous studies splitting $CO_2$ to CNT by constant current electrolysis in molten carbonate generally were studied at moderate current densities ranging from 0.03 to $0.4 A/cm^2$. Figure S2 presents the SEM of cleaned cathode product synthesized at a higher electrolysis current density of $0.6 A/cm^2$ for 30-min conducted in a molten electrolyte (750°C molten $Li_2CO_3$ containing 2wt% dissolved $Li_2O$) using a Muntz brass cathode and Nichrome C anode. As with electrolyses at lower current densities, based on the product mass, the electrolysis occurred at >> 90% coulombic efficiency. As seen the product is highly uniform and are straight; based on the product's mass the product is formed at over 98% coulombic efficiency. The CNT product purity is 97% purity and the CNTs are ~150μm long. Similar syntheses in which the 2 wt% $Li_2O$ additive was replaced by 10wt% dehydrated borax, and in which the Muntz Brass Cathode was replaced by a Monel Cathode, or in which the NiCr anode was replaced by an Inconel 718, or SS304 or C-264 produced similar results.

The Raman spectrum of the CNTs synthesized at $0.6 A/cm^2$ in a 750°C molten $Li_2CO_3$ containing 2wt% dissolved $Li_2O$ is shown in Figure S3. The Raman spectrum exhibits two sharp peaks observed around 1350 and $1580 cm^{-1}$, which correspond to the disorder-induced mode (D band) and the high frequency E2g first order mode (G band), respectively and an additional peak, the 2D band, at 2700cm$^{-1}$. The intensity ratio between D band and G band ($I_D/I_G$) and 2D band and G band ($I_{2D}/I_G$) are useful parameters to evaluate the number of defects and degree of graphitization. In Figure S3 the measured respective $I_D/I_G$ and $I_{2D}/I_G$ of 0.4 and 0.7 are characteristic of high quality (low defect) multiwalled CNT.

XRD of the CNTs synthesized at 0.6A/cm$^2$ in a 750°C molten $Li_2CO_3$ containing 2wt% dissolved $Li_2O$ are shown in Figure S4 and by comparison to the library XRD included in the lower portion of the figure are shown to consist of a dominant graphitic structure with contributions from $Fe_3C$ and $LiNiO_2$.

TEM and EDX of the CNTs synthesized at 0.6 A cm$^{-2}$ in a 750°C molten $Li_2CO_3$ containing 2 wt% dissolved $Li_2O$ are shown in Figures S5 and S6. Figure S5 includes an image spot of the carbon nanotube core that contains metal. In S5 and S6 the element carbon comprising the CNT walls is evident on the outer portions of the CNT and the metals in S5 (iron) is evident within. We have recently probed iron carbide within CNTs as a mechanism to facilitate CNTs [33].

For the electrochemical growth of HCNTs by $CO_2$ electrolysis, in the absence of

(i) high electrolysis current density growth,

(ii) high temperature electrolysis growth (such as 770°C or higher as seen in Figure 3),

(iii) a sp$^3$ defect inducing agents, such as added oxide, and/or

(iv) a controlled concentration of iron added to the electrolyte or cathode surface

curved & bent CNTs can be formed, but as shown in Figure S7, panel A, are produced without a regular, repeating helical morphology.

The conditions investigated in this study are used to demonstrate high yield formation HCNT, HCNP or HCNF, but other molten electrolyte conditions can synthesize such materials, albeit at low yield as also shown in Figure S7. For example, in a pure $Li_2CO_3$ electrolyte (without $Li_2O$ or $Fe_2O_3$ additives) at a relatively high current density of 0.6A/cm$^2$, a smaller fraction of ~20% HCNTs and HCNPs are produced along with a majority of curled CNTs as shown in Figure S7, panel B. As shown in Panels C and D, HCNTs and HCNPs can be synthesized in certain ternary electrolytes, even at lower current density and in the absence of $Fe_2O_3$, albeit at lower yield. Specifically, the product of a molten carbonate electrolysis containing 20wt% $Na_2CO_3$ and 80wt% $Li_2CO_3$ and an additional additive of 8wt% dehydrated borax was used to split $CO_2$ at a low current density of $0.2 A/cm^2$. While as shown in Panel C the major product was curled CNTs, also as shown in Panel D a significant co-product was HCNTs and HCNPs.

This data section provides baseline and comparative experimental data for non-helical analogues to a recently discovered, novel inexpensive electrosynthesis of high purity helical carbon nanomaterials.

**Acknowledgements**
We are grateful to C2CNT for support of this research.

**Declaration of Interests**
The authors declare that they have no known competing financial interests or personal relationships that could have appeared to influence the work reported in this paper.
.

**References**

[1] N. Tang, J. Wen, Y. Zhang, F. Liu, K. Lin, Y. Du, Helical Carbon Nanotubes: Catalytic Particle Size-Dependent Growth and Magnetic Properties, ACS Nano 4(1), (2010) 241.

[2] R. Gao, Z. L. Wang, S. Fan, Kinetically Controlled Growth of Helical and Zigzag Shapes of Carbon Nanotubes, J. Phys. Chem. B 104 (2000) 1227.

[3] Y. Qin, Z. Zhang, Z. Cui, Helical carbon nanofibers with a symmetric growth mode, Carbon 42(10) (2004) 1917.

[4] Y. Suda, Chemical Vapor Deposition of Helical Carbon Nanofibers, Chemical Vapor Deposition for Nanotechnology (2018) *intechopen. 81676,*

[5] V. Bajpai, L. Dai, T. Ohashi, Large-Scale Synthesis of Perpendicularly Aligned Helical Carbon Nanotubes, J. Am. Chem. Soc. 126(16) (2004) 5070.

[6] K. Tak, M. Lu, D. Hui, Coiled carbon nanotubes: Synthesis and their potential applications in advanced composite structures, Composites Part B: Engineering *37*(6) (2006) 437.

[7] M. Zhang, J. Li Carbon nanotube in different shapes, Materials Today 12(6) (2009) 12.

[8] W. Wang, K Yang, J. Galliard, P. R. Bandaru, Q. M. Rao Rational Synthesis of Helically Coiled Carbon Nanowires and Nanotubes through the Use of Tin and Indium Catalysts, Advanced Materials 20(1) (2008) 179.

[9] Q. Zhang, M. Zhao, D. Tang, F. Li, J. Huang, B. Liu, W. Zhu, Y. Zhang, F. Wei, Carbon-Nanotube-Array Double Helix. Angew. Chem. Int. Ed. 49 (2010) 3642.

[10] N. Komatsu, F. A. Wang, Comprehensive Review on Separation Methods and Techniques for Single-Walled Carbon Nanotubes, Materials (Basel) 3(7) (2010) 3818.

[11] S. Itoh, S. Ihara, J. Kitakami, Toroidal Form of Carbon C360, Phys. Rev. B 47 (1993) 1703.

[12] S. Amelinckx, X. B. Zhang, D. Bernaerts, X. F. Zhang, V. Ivanov, J. B. Nagy, A Formation Mechanism for Catalytically Grown Helix-Shaped Graphite Nanotubes, Science 265 (1994) 635.

[13] X. B. Zhang, X. F. Zhang, D. Bernaerts, G. Van Tendeloo, S. Amelinckx, J. V. Landuyt, V. Vanov, J. B. Nagy, P. Lambin, A. A, Lucas. The Texture of Catalytically Grown Coil-Shaped Carbon Nanotubules, Euro. Phys. Lett. 27 (1994) 141.

[14] V. Khanna, B. R. Bakshi, L. J. Lee, Carbon Nanofiber Production: Life Cycle Energy Consumption and Environmental Impact, J. Ind. Ecology 12 (2008*)* 394.

[15]   J. Houghton, Y. Ding, D. Griggs, M. Noguer, P. van der Linden, X. Dai, K. Maskell, C. Johnson. Climate Change 2001: The Scientific Basis, Cambridge University Press (2001).

[16]   K. Trenberth, Uncertainty in Hurricanes and Global Warming, Science *308* (2005) 1753.

[17]   Thomas, C, A. Cameron, R. Green, M. Bakkenes, L. Beaumont, Y. Collingham, B. Erasmus, M. Ferreira de Siqueira, A. Grainger, L. Hanna, L. Hughes, B. Huntley, A. van Jaarsveld, G. Midgley, L. Miles, M. Ortega-Huerta, A. Townsend Peterson, O. Phillips, S. Williams, Extinction risk from Climate Change, Nature 427 (2004) 145.

[18]   National Research Council, Advancing the Science of Climate Change, Washington, DC: The National Academies Press (2010).

[19]   B. E. Walling, Z. Kuang, Y. Hao, D. Estrada, J. D. Wood, F. Lian, L. A. Miller, A. B. Shah, J. L. Jeffries, R. T. Haasch, Helical Carbon Nanotubes Enhance the Early Immune Response and Inhibit Macrophage-Mediated Phagocytosis of *Pseudomonas aeruginosa, PLOS ONE* 8(11) (2013) e80283.

[20]   J. Ren, F. F. Li, J. Lau, L. González-Urbina, S. Licht, One-Pot Synthesis of Carbon Nanofibers from $CO_2$, Nano Lett. 15(9) (2015) 6142.

[21]   J. Ren, J. Lau, M. Lefler, S. Licht, The Minimum Electrolytic Energy Needed to Convert Carbon Dioxide to Carbon by Electrolysis in Carbonate Melts, J. Phys. Chem. C 119(30) (2015) 23342.

[22]   M. Johnson, J. Ren, M. Lefler, G. Licht, J. Vicini, X. Liu, S. Licht, Carbon Nanotube Wools Made Directly from CO2 by Molten Electrolysis: Value Driven Pathways to Carbon Dioxide Greenhouse Gas Mitigation, Mater. Today Energy 5 (2017) 230.

[23]   M. Johnson, J. Ren, M. Lefler, G. Licht, J. Vicini, X. Liu, S. Licht, Data on SEM, TEM and Raman Spectra of Doped, and Wool Carbon Nanotubes Made Directly from $CO_2$ by Molten Electrolysis, Data Br. 14 (2017) 592.

[24]   J. Ren, M. Johnson, R. Singhal, S. Licht, Transformation of the Greenhouse Gas CO2 by Molten Electrolysis into a Wide Controlled Selection of Carbon Nanotubes, J. CO2 Util. 18 (2017) 335.

[25]   S. Licht, A. Douglas, J. Ren, R. Carter, M. Lefler, C. L. Pint, Carbon Nanotubes Produced from Ambient Carbon Dioxide for Environmentally Sustainable Lithium-Ion and Sodium-Ion Battery Anodes, ACS Cent. Sci. 2(3) (2016) 162.

[26]   J. Ren, S. Licht, Tracking Airborne $CO_2$ Mitigation and Low Cost Transformation into Valuable Carbon Nanotubes. Sci. Rep. (6) (2016) 27760.

[27] X. Liu, J. Ren, G. Licht, X. Wang, Licht, S. Carbon Nano-onions Made Directly from $CO_2$ by Molten Electrolysis for Greenhouse Gas Mitigation, Adv. Sustainable Materials (2019) 1900056.

[28] S. Licht, X. Liu, G. Licht, X. Wang, A. Swesi, Y. Chan, Amplified $CO_2$ reduction of greenhouse gas emissions with C2CNT carbon nanotube composites, Materials Today Sustainability 6 (2019) 100023.

[29] X. Liu, X. Wang, G. Licht, B. Wang, S. Licht, Exploration of alkali cation variation on the synthesis of carbon nanotubes by electrolysis of $CO_2$ in molten carbonates, J. $CO_2$ Utilization 18 (2019) 378.

[30] J. Ren, A. Yu, P. Penge, M. Lefler, F.-F. Li, S. Licht, Recent Advances in Solar Thermal Electrochemical Process (STEP) for Carbon Neutral Products and High Value Nanocarbons, *Accounts of Chemical Research* 52 (2019) 3177.

[31] X. Liu, X. Wang, G. Licht, S. Licht, Transformation of the greenhouse gas carbon dioxide to graphene, J. $CO_2$ Utilization 36 (2020) 288.

[32] X. Wang, X. Liu, G. Licht, S. Licht, Calcium metaborate induced thin walled carbon nanotube syntheses from $CO_2$ by molten carbonate electrolysis, Scientific Reports 10 (2021) 15416.

[33] X. Wang, X. Liu, S. Licht, Magnetic carbon nanotubes: Carbide nucleated electrochemical growth of ferromagnetic CNTs from $CO_2$, J. $CO_2$ Utilization 40 (2020) 10128H.

[34] X. Wang, G. Licht, X. Liu, S. Licht, One pot facile transformation of $CO_2$ to an unusual 3-D nano-scaffold morphology of carbon, Scientific Reports 10 (2020) 21518.

[35] X. Wang, G. Licht, S. Licht, Green and scalable separation and purification of carbon materials, Separation and Purification Technology 255 (2021) 117719.

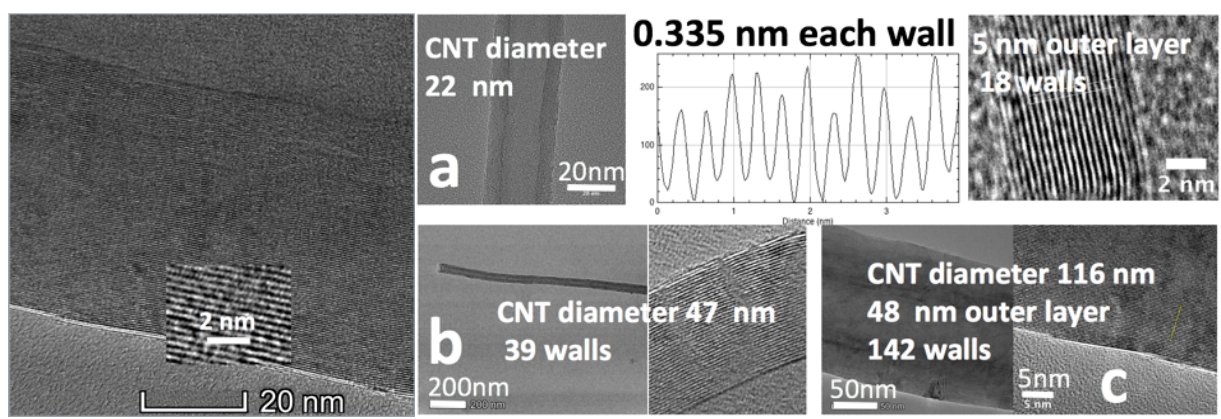

**Figure S1.** TEM (Transmission electron microscopy) of CNT walls in molten carbonate synthesized CNTs. The synthesis is by electrolysis in 770°C Li$_2$CO$_3$, with a 5cm$^2$ coiled copper wire. Left: an expanded view of the carbon nanotube product after 90-min synthesis. The synthesis produces a pure CNT product whose diameter increases with electrolysis time. TEM of the synthesis product subsequent to a: 15, b: 30 or c: 90-min electrolysis.

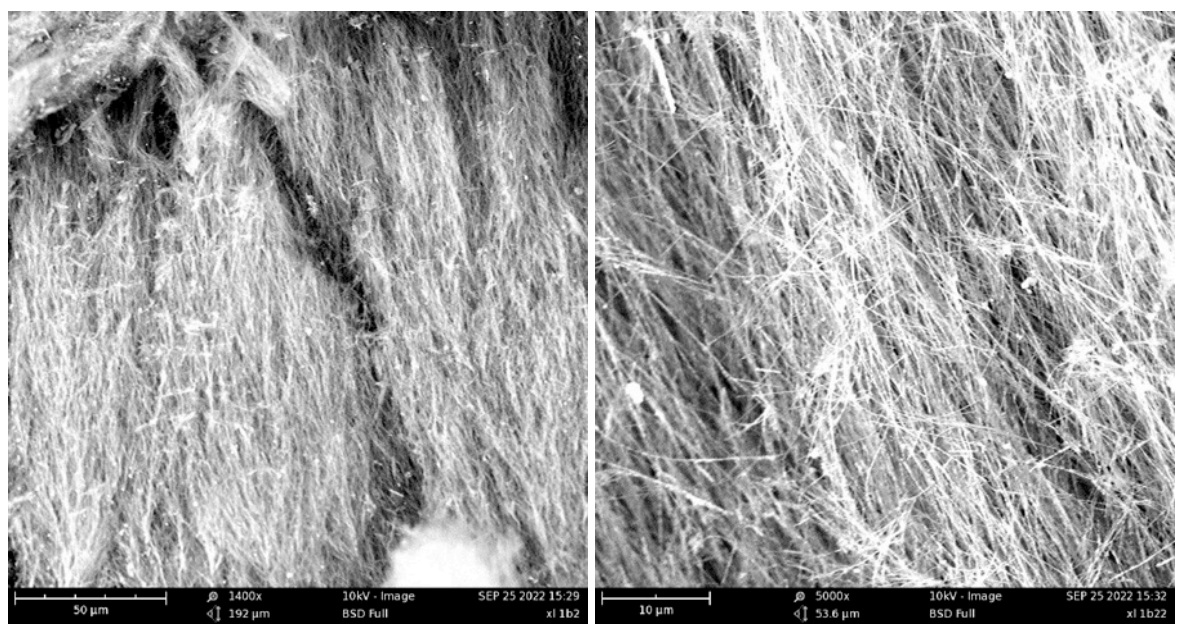

**Figure S2.** SEM of conventional high current density C2CNT CO$_2$ electrolysis. The washed product is collected from the cathode subsequent to a 30 minute, high current density electrolysis (0.6A/cm$^2$, compare to typical 0.03 to 0.3A/cm$^2$ electrolyses) in 750°C Li$_2$CO$_3$ with 2wt% Li$_2$O). The cathode is a planar 27cm$^2$ Muntz Brass vertically separated 1cm from a 27cm$^2$ Nichrome C planar anode.

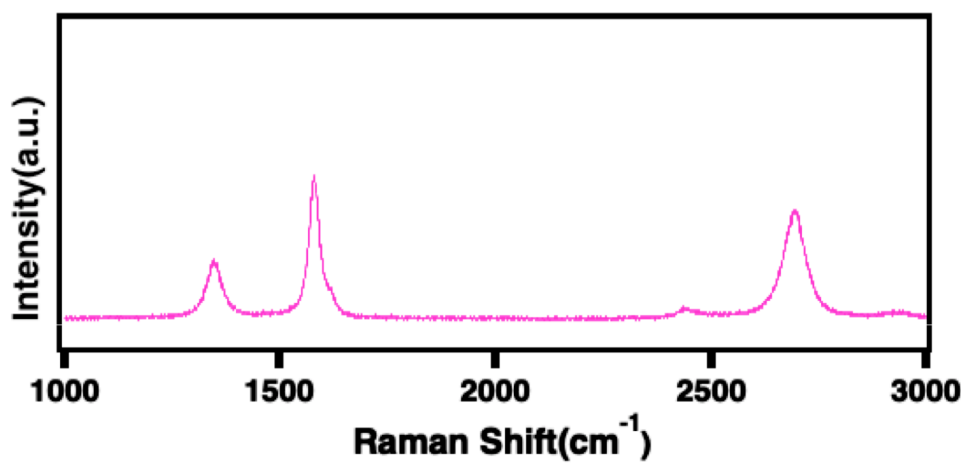

**Figure S3.** Raman Spectrum of the carbon nanotubes described in Figure S2 (synthesized by electrolysis at 0.6 A/cm$^2$ in 750°C Li$_2$CO$_3$ with 2 wt% Li$_2$O).

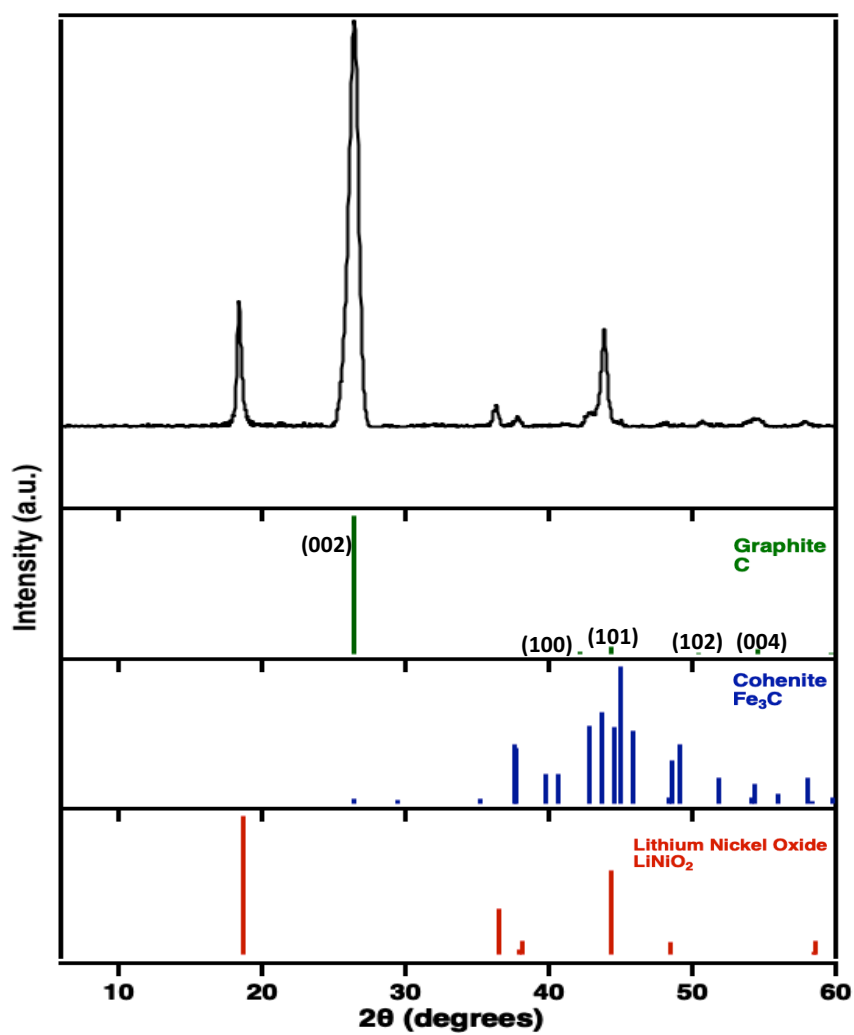

**Figure S4. Top:** XRD of the carbon nanotubes described in Figure S2 (synthesized by electrolysis at 0.6 A cm$^{-2}$ in 750°C Li$_2$CO$_3$ with 2 wt% Li$_2$O). Middle and bottom: Library XRF of graphite, Fe$_3$C and LiNiO$_2$.

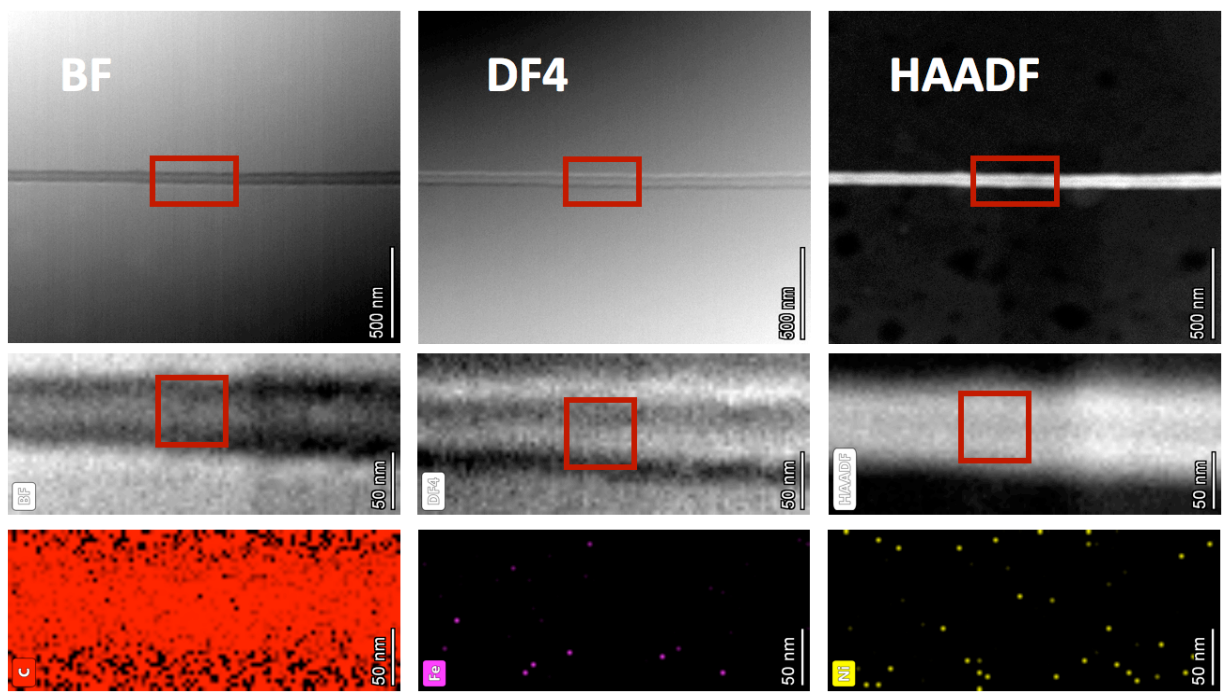

**Figure S5. Top:** TEM, spot 1, from the carbon nanotube sample described in Figure S2 including BF (bright-field), DF (dark-field) and HAADF (high-angle annular dark-field) imaging. **Bottom:** EDX spectra of this spot (red = carbon; purple = iron; yellow = nickel).

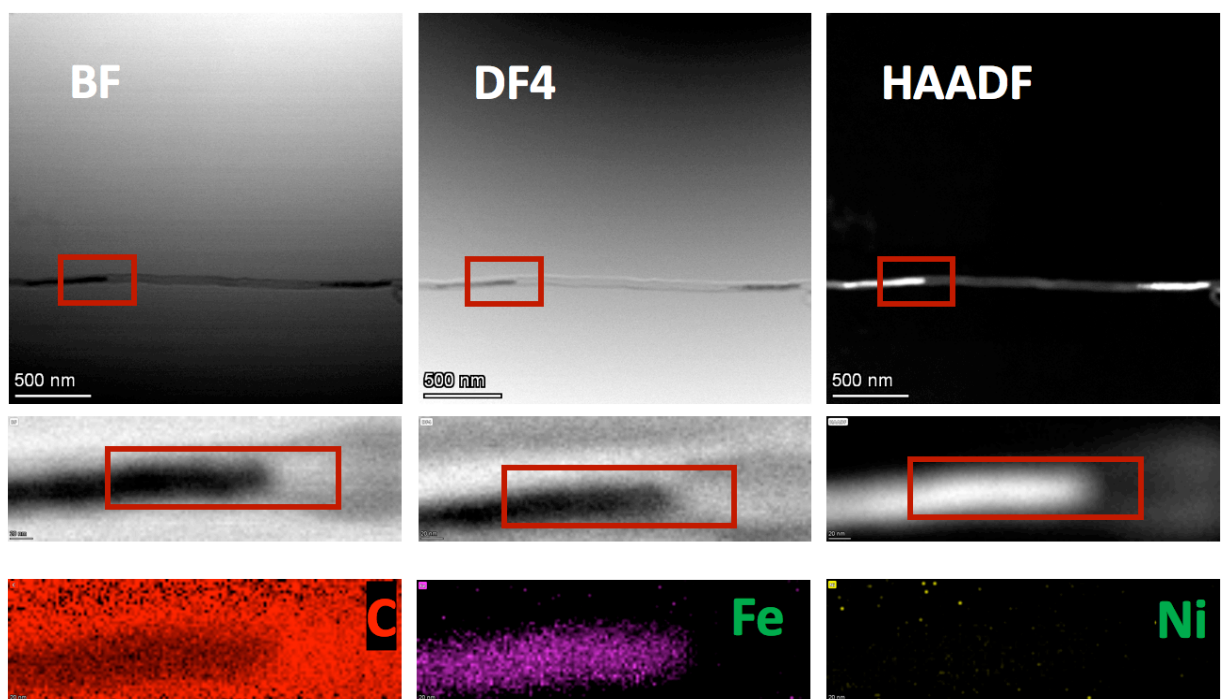

**Figure S6. Top:** TEM, spot 2, from the carbon nanotube sample described in Figure S2 including BF (bright-field), DF (dark-field) and HAADF (high-angle annular dark-field) imaging. **Bottom:** EDX spectra of this spot (red = carbon; purple = iron; yellow = nickel).

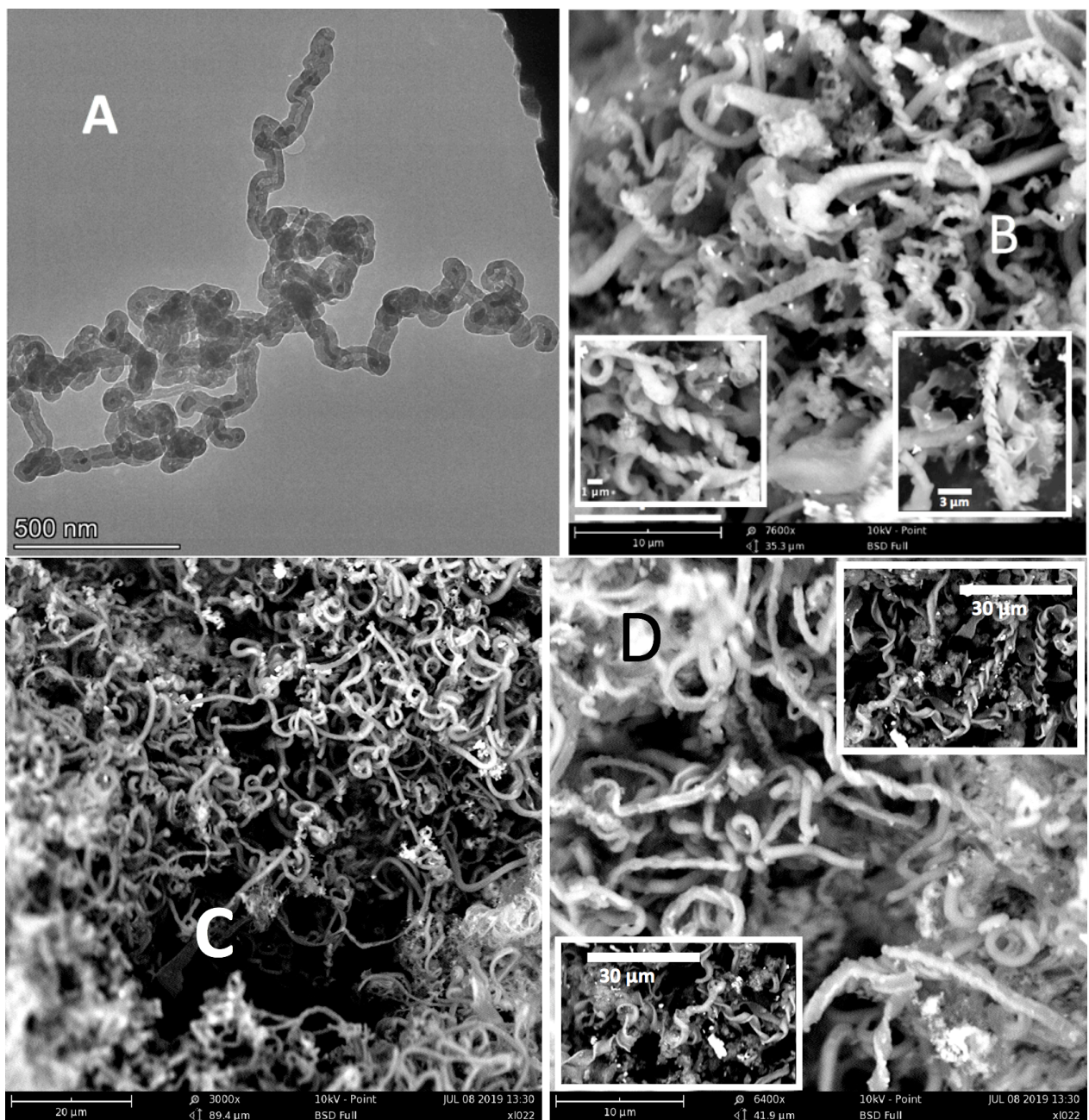

**Figure S7.** Panel A TEM, Rather than straight CNTs as in Figure S2, or HCNTs as in Figures 6 and 7 of the main paper "The green synthesis of exceptional braided, helical carbon nanotubes and nano spiral platelets made directly from $CO_2$.", the product of uncontrolled electrolytic $CO_2$ splitting in molten carbonates can lead to bent and deformed CNTs. Panels B-D SEM. Panel B, with only partial control of higher electrolysis current densities and no defect inducing agents, some HCNTs and HCNPs can be produced, albeit at lower yields. Shown is the product of a 0.5A/cm$^2$ 2 hour electrolysis in pure $Li_2CO_3$ (without additives) using 30cm$^2$ area planar Nichrome C anode and Muntz Brass cathode. Panels C and D, show HCNTs and HCNPs can be synthesized in certain ternary electrolytes, even at lower current density and in the absence of $Fe_2O_3$, albeit at lower yield. Specifically, the product of a molten carbonate electrolysis containing 20wt% $Na_2CO_3$ and 80wt% $Li_2CO_3$ and an additional additive of 8wt % dehydrated borax was used to split $CO_2$ at a low current density of 0.2A/cm$^2$.